\documentclass[10pt,letter,aps,prb,showpacs,superscriptaddress,twocolumn,
floatfix]{revtex4}
\usepackage{amsmath}
\usepackage{citesort}
\usepackage{graphicx}
\begin{document}
\title{Optical conductivity in non-equilibrium $d$-wave superconductors}
\author{J.P. Carbotte}
\affiliation{Department of Physics and Astronomy, McMaster University,\\
Hamilton, Ontario, L8S 4M1 Canada}
\author{E. Schachinger}
\email{schachinger@itp.tu-graz.ac.at}
\homepage{www.itp.tu-graz.ac.at/~ewald}
\affiliation{Institute of Theoretical and Computational Physics\\
Graz University of Technology, A-8010 Graz, Austria}
\date{\today}
\begin{abstract}
  We consider the optical conductivity of a $d$-wave BCS superconductor
  in the presence of a non-equilibrium distribution of excess
  quasiparticles. Two different simplified models used in the past
  for the $s$-wave case are considered and results compared. In the
  $T^\ast$-model of Parker the excess quasiparticles are assumed
  to be in a thermal distribution at some temperature $T^\ast$
  larger than the equilibrium sample temperature. In the
  $\mu^\ast$-model of Owen and Scalapino a chemical potential
  is introduced to accommodate the excess quasiparticles. Some of the
  results obtained are specific to the model, most are
  qualitatively similar in both.
\end{abstract}
\maketitle
\newpage
\section{Introduction}

The study of non-equilibrium superconductivity in conventional
$s$-wave superconductors has a long history. Several comprehensive
reviews exist.\cite{rev1,rev2,rev3} Much less has been done for
unconventional order parameters including the high $T_c$ oxides. The
cuprates are particularly interesting since there is now a
consensus that its energy gap has $d$-wave\cite{hardy,shen,%
wollm,tsuei} rather than $s$-wave symmetry. This difference is
significant particularly since it implies the existence of
gap nodes on the Fermi surface and hence the existence of
quasiparticle states of low energies.

One might expect that quasiparticle excitations created at rather
high energies through application of a laser pulse to the
equilibrium sample or by some other means, would rapidly relax
towards the lowest energy states available and accumulate in
the nodal regions of the gap on the Fermi surface where the
density of excited states goes linearly in energy. The fast
relaxation processes can proceed through the electron-electron
interaction which is strong in the cuprates and also through
coupling to the phonons.\cite{allen} The phonons are a separate
system of excitations which can share in the excess energy
provided by the excitation source. This is to be contrasted to
what happens in
formulations of highly correlated systems in terms of
boson exchange processes,\cite{schach4,schach5,schach7,schach6,%
schach8,schach9} such as the exchange of spin
fluctuations\cite{pines2} in the case of the Nearly Antiferromagnetic
Fermi Liquid (NAFFL) model. While such
exchange processes can equilibrate the energy, they do not
take it out of the electronic system. Another feature of
$d$- as opposed to $s$-wave non-equilibrium superconductivity
is that, as has been argued by Feenstra {\it et al.},%
\cite{feenstra} the final stage of
equilibration which proceeds through recombination of nodal
quasiparticles into Cooper pairs, may take a much larger time than in
$s$-wave because of restrictions on the kinematics.
New fast laser pulse techniques\cite{kabanov,carr,%
woerner} have helped greatly in the study of
non-equilibrium state in cuprates. These techniques have recently
been used in lead\cite{carr} to measure the conductivity
$\sigma(T,\omega)$ under non-equilibrium conditions as a function
of frequency in the entire infrared range.
Similar experiments in the cuprates should be
possible and we hope this paper will stimulate such work. Here we provide a
calculation of non-equilibrium effects just before recombination
on the optical conductivity of a $d$-wave superconductor. We also
consider explicitely the microwave region where analytic results
can be obtained.

Very recently Nicol and Carbotte\cite{nicol} studied the effect
of a non-equilibrium distribution of quasiparticles on the gap
in a BCS $d$-wave superconductor. In this work
two well established models coming from the literature on conventional
non-equilibrium superconductors are employed. These are the
$\mu^\ast$-model by Owen and Scalapino\cite{owen} and the $T^\ast$-model of
Parker.\cite{parker1,parker2} These models, while very simple,
have been useful in conceptualizing what can happen,
in planning, and interpreting experiments. In some sense
they represent two extreme limits between which the
true distribution might be expected to fall. The $\mu^\ast$-%
model assumes that the excess quasiparticles rapidly
accumulate in the lowest quasiparticle energy states available
which is around the gap $\Delta$ in $s$-wave
but is around zero in $d$-wave
where the density of states shows a linear in energy
dependence. A finite
chemical potential $\mu^\ast$ (for the quasiparticles) is used
to describe the occupation. The other extreme is to assume that the
excess quasiparticles equilibrate at some new temperature
$T^\ast$ larger than the sample temperature $T$. In both
cases the system is not in thermal equilibrium because the
measurement is to be made before the final recombination
process into Cooper pairs proceed. This is assumed to
be the largest time scale in the relaxation process.

There are significant differences in
behavior for $d$-wave as compared to $s$-wave. For example
in terms of the excess quasiparticle density $n$ (to be defined
more precisely later) the reduction in the gap $\delta\Delta/\Delta$
is given by $-2n$ for both, $T^\ast$ and $\mu^\ast$ models in
$s$-wave, but for $d$-wave it goes like $-(4\sqrt{2}/3)%
\sqrt{n^3}$ for the $\mu^\ast$-model and
$-(32/\pi^3)\sqrt{(3n)^3}$ in the $T^\ast$ model. These
fundamentally different dependences on $n$, linear in $s$-wave,
and to a 3/2 power in $d$-wave can be simply understood.
The presence of excess quasiparticles blocks states around the
Fermi surface which would otherwise be available to form the
condensate in a variational sense. In $d$-wave,
the excitations are assumed to gather around the nodal regions.
In these regions the gap is small and contributes
little to the condensation energy so that blocking such states
is less effective than in the $s$-wave case where the gap is
everywhere finite and equal to $\Delta$.

In their paper Nicol and Carbotte\cite{nicol} consider explicitly
the case of tunneling in a normal-insulator-superconductor (N-I-S)
junction with a non-equilibrium distribution on the superconducting
side. They also address the case of pump probe experiments which
claim to measure the temperature dependence of the excess
quasiparticles $n(T)$ before final recombination into Cooper pairs.
In the present paper we extend the work to the optical
conductivity. Such calculations exist in the $s$-wave case\cite{chang}
based on an appropriate generalization of the dirty limit
Mattis-Bardeen formula\cite{mattis} for the optical conductivity.
Here we consider the $d$-wave case\cite{schur}
and treat explicitly both, the $T^\ast$ and the $\mu^\ast$
model and  compare results obtained for the same value
of excess quasiparticles $n$. Since in many aspects
both models give the same qualitative physics, although
there are some important differences which we will note, we can
expect that more
realistic distributions would not be so different.

In Section II we introduce some elements of the theory of the optical
conductivity $\sigma(\omega,T)$ as a function of frequency
$\omega$ at
temperature $T$ suitably generalized for the non-equilibrium case.
In the $T^\ast$-model no formal changes to the formulas are
needed, only the temperature is to be interpreted differently. In the
$\mu^\ast$-model all necessary changes to the mathematical formulas
can be incorporated
through a change of the thermal factors. In Section III we present
exact numerical results for the frequency dependence of the
optical conductivity (real and imaginary part) as well as for
the reflectance and the normalized reflectance difference due
to an excess distribution ($n$) of non-equilibrium quasiparticles
which is the quantity that is directly measurable.
Results for the $T^\ast$ and $\mu^\ast$-model for the same $n$
are compared. Section IV contains a number of analytic results for a
regime in which the temperature is larger than the impurity scattering rate
i.e. the weak scattering limit. Comparison with exact
numerical results is also given. Dependences on the excess
quasiparticle density $n$ are made explicit and helps
physical understanding. A short conclusion is found in Section V.
An Appendix contains some mathematical details.

\section{Theory}

The non-equilibrium distribution in either $T^\ast$ or $\mu^\ast$-%
model does not change the symmetry of the gap. It does reduce its
magnitude. In addition, the thermal factors which enter directly
the infrared conductivity formula are also changed. These changes
are modeled in one case by the introduction of a chemical
potential $\mu^\ast$ and in the other by simply changing the
temperature from $T$, the temperature of the sample before
application of a laser pulse, to $T^\ast$ the non-equilibrium
quasiparticle temperature before final recombination. For simplicity,
we will assume here
that the initial sample temperature $T$ is sufficiently low that
it can be considered to be zero. In the actual numerical work
it is taken to be $3\,$K for convenience with a gap of $24\,$meV.
For a BCS $d$-wave superconductor this choice corresponds to a
$T_c$ of $130\,$K. If we further assume that the density of
excess quasiparticles $n$ is small, we can obtain analytic results
for the change in gap and for the corresponding value of
$\mu^\ast$ and/or $T^\ast$. The necessary results are to be found
in the paper by Nicol and Carbotte.\cite{nicol} In the $T^\ast$-%
model $n = (\pi^2/12)(T^\ast/\Delta_0)^2$ and $(\delta\Delta%
/\Delta_0) = -(32/\pi^3)\sqrt{(3n)^2}$, while in the $\mu^\ast$-%
model $\mu^\ast/\Delta_0 = \sqrt{2n}$ and
$\delta\Delta/\Delta_0 = -(4\sqrt{2}/3)\sqrt{n^3}$. [From now on
we will use $\Delta$ or $\Delta(n)$ for the gap with a finite
$n$ and $\Delta_0$ for its equilibrium value when $n=0$, i.e.
$\Delta_0=\Delta(n=0)$.]
These
relationships can be used directly in our conductivity calculations.
Some explanation is needed. The above results were derived for
a $d$-wave BCS superconductor in the pure limit. To get
meaningful conductivity results it is necessary to formally
include some scattering mechanism. Here we will simply treat
elastic impurity scattering. When this is done the impurities
also formally enter the gap equation and in principle one
should include these modifications along with those brought about
by the non-equilibrium distribution of quasiparticles in the
formula for the conductivity. Here,
for simplicity, we will not do this. The impurities do reduce the
size of the $d$-wave gap amplitude but we will assume that this
is already included in our initial choice of this parameter and
simply treat the changes brought about by $n$ based on the pure
case. Assuming that the impurity concentration is very small as
we wish to do here, this is sufficient. There will also be
a corresponding effect of impurities on the chemical potential
$\mu^\ast$ but this is also neglected in the limit of small
impurity scattering. With this explanation
we can proceed with the calculation of the optical conductivity.

It is conventional, and we will follow this definition here, to
measure the excess quasiparticle density $n$ in units of $4N(0)\Delta_0$
where the factor 4 is from spin degeneracy and restricting the
defining integral to positive energies only. $N(0)$ is the
normal state electronic density of states at the Fermi surface.
In these units
\begin{subequations}
  \label{eq:1}
  \begin{equation}
  \label{eq:1a}
  n = \frac{1}{\Delta_0}\left\langle\int\limits_0^\infty\!
    d\epsilon_{\bf k}\,\left[f_T(E_{\bf k}-\mu^\ast)-
      f_T(E_{\bf k})\right]\right\rangle_\theta
  \end{equation}
  for the $\mu^\ast$-model and
  \begin{equation}
    \label{eq:1b}
   n = \frac{1}{\Delta_0}\left\langle\int\limits_0^\infty\!
    d\epsilon_{\bf k}\,\left[f_{T^\ast}(E^\ast_{\bf k})-
      f_T(E_{\bf k})\right]\right\rangle_\theta    
  \end{equation}
\end{subequations}
in the $T^\ast$-model. Here $f_T(\epsilon)$ is the usual Fermi
Dirac distribution $f_T(\epsilon) = [1+\exp(\beta\epsilon)]^{-1}$
with $\beta = 1/k_B T$ where $k_B$ is the Boltzmann factor
which we set to one from here on. In
Eqs.~(\ref{eq:1}) the brackets $\langle\cdots%
\rangle_\theta$ imply an angular average over the polar angle
$\theta$ which gives the position on the Fermi surface in the
two dimensional CuO$_2$ Brillouin zone. In terms of this angle
the gap $\Delta(\theta) = \Delta\cos(2\theta)$.
The quasiparticle energies
$E_{\bf k} = \sqrt{\epsilon^2_{\bf k}+\Delta^2(T,\theta)}$ where
we make explicit the temperature dependence of the
gap which enters the $T^\ast$-model. Here {\bf k} is momentum
and $\epsilon_{\bf k}$ is the normal state band energy.
Consistent with our treatment of the gap equation the clean
limit is used in writing down Eqs. (\ref{eq:1}).

The formula for the optical conductivity in the equilibrium
case takes on the form:\cite{schur,hirschf,mars1,mars2}
\begin{widetext}
\begin{eqnarray}
  \label{eq:2}
  \sigma(T,\nu) &=& \frac{\Omega^2_p}{4\pi}\frac{i}{\nu}
     \left\langle
      \int\limits_0^\infty\!d\omega\,\tanh\left(\frac{\beta\omega}{2}
        \right)\left[
        J(\omega,\nu)- J(-\omega,\nu)
      \right]\right\rangle_\theta\nonumber\\
  &=& \frac{\Omega^2_p}{4\pi}\sigma'(T,\nu), 
\end{eqnarray}
where $\Omega_p$ is the plasma frequency and the function $J(\omega,\nu)$
is
\begin{eqnarray}
  2J(\omega,\nu) &=& \frac{1}{E_1+E_2}\left[1-N(\omega)N(\omega+\nu)
    -P(\omega)P(\omega+\nu)\right]\nonumber\\
  &&+\frac{1}{E^\ast_1-E_2}\left[1+N^\ast(\omega)N(\omega+\nu)
    +P^\ast(\omega)P(\omega+\nu)\right],
  \label{eq:3}
\end{eqnarray}
\end{widetext}
with $E^\ast_1(\omega)$, $N^\ast(\omega)$, and $P^\ast(\omega)$ 
the complex conjugate of $E_1(\omega)$, $N(\omega)$, and $P(\omega)$,
respectively. In Eq.~(\ref{eq:2})
\begin{eqnarray*}
  E_1(\omega) &=& \sqrt{\tilde{\omega}^2(\omega+i0^+)-
    \tilde{\Delta}^2(\omega+i0^+)},\\
  \qquad E_2(\omega,\nu) &=& E_1(\omega+\nu),
\end{eqnarray*}
and
\[
  N(\omega) = \frac{\tilde{\omega}(\omega+i0^+)}{E_1(\omega)},
  \qquad P(\omega) = \frac{\tilde{\Delta}(\omega+i0^+)}{E_1(\omega)}.
\]
In these equations $\tilde{\omega}(\omega+i0^+)$ is the renormalized
frequency which includes the impurities and in the pairing energy
$\tilde{\Delta}(\omega+i0^+)$ the angular dependence on $\theta$
has been suppressed. $\sigma'(T,\nu)$ is the optical conductivity
as it is calculated within our numerical programs. It is given in
units of meV$^{-1}$. If $\Omega_p$ is given in meV then
$\sigma(T,\nu)$ is given in units of meV which can easily be
transformed to SI units using the relation that one $\Omega^{-1}\,
{\rm m}^{-1}$ corresponds to $5.916\times 10^{-3}\,$meV. Finally,
in BCS $\tilde{\Delta}(\omega+i0^+)$ is independent of $\omega$.

 To treat a non-equilibrium system a chemical potential $\mu^\ast$
for the excess quasiparticles is introduced in the grand canonical
ensemble average and this means that
the thermal factor in Eq.~(\ref{eq:2}) needs to be replaced by:
\begin{eqnarray*}
  \label{eq:4}
  \tanh\left(\frac{\beta\omega}{2}\right)&\longrightarrow&
  F(\omega)\equiv\left[-f_T(\omega+\mu^\ast)\right.\\
  &&\left.-f_T(\omega-\mu^\ast)+1\right]
\end{eqnarray*}
while in the $T^\ast$-model there are no modifications except to
change the temperature $T$ to $T^\ast$.
In Fig.~\ref{fig:1} we compare the two factors:
  \begin{figure}[t]
    \includegraphics[width=8cm]{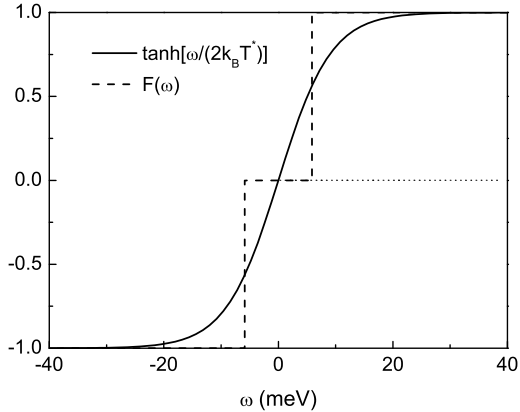}  
    \caption{The thermal factor $\tanh(\omega/2k_BT^\ast)$ which
    enters the optical conductivity (Eq.~(\protect{\ref{eq:2}}))
    in the $T^\ast$-model (solid curve) compared with the equivalent
    function $F(\omega)\equiv\left[-\theta(\omega+\mu^\ast)-
    \theta(\omega-\mu^\ast)+1\right]$ which enters the $\mu^\ast$
    model (dashed curve). As shown in the text for a common value
    of excess quasiparticles $n$, $2T^\ast = (2\sqrt{6}/\pi)\mu^\ast%
    \simeq 1.56\mu^\ast$. This relation is built into the figure. Here
    $T^\ast = 53\,$K and $\mu^\ast = 5.88\,$ have been chosen.
    }
    \label{fig:1}
  \end{figure}
$F(\omega)$ and $\tanh(\omega/2 k_B T^\ast)$ with $\mu^\ast$
related to $T^\ast$ by $T^\ast \simeq 1.56\,\mu^\ast$. We see that both
are antisymmetric with respect to $\omega=0$ and that $F(\omega)$
depletes the region around $\omega=0$ more than does the hyperbolic
tangent. These differences  will reflex themselves in the conductivity.
We begin by giving results of our numerical evaluation of Eq.~(\ref{eq:2}).
Before proceeding, it is necessary to specify how
impurities are to be included in the theory. The impurities
renormalize the frequencies to $\tilde{\omega}(\omega+i0^+)$ from
$\omega$ according to the equation\cite{hirschf1,schach1}
\begin{equation}
  \label{eq:5}
  \tilde{\omega}(\omega+i0^+) = \omega+i\pi\Gamma^+
    \frac{\left\langle N(\tilde{\omega})\right\rangle_\theta}
    {c^2+\left\langle N(\tilde{\omega})\right\rangle^2_\theta},
\end{equation}
where we have introduced the impurity scattering in a $T$-matrix
formalism, $c=1/\left(2\pi N(0)V_{imp}\right)$.
Here $V_{imp}$ is the strength of
the impurity potential and $N(0)$ the electronic density of
states at the Fermi surface in the normal state which is taken
to be constant, i.e. energy independent in the range of
energies relevant to superconductivity. In Eq.~\eqref{eq:5} the
real part of $\left\langle N(\tilde{\omega})\right\rangle_\theta$
is the quasiparticle density of states. The parameter
$\pi\Gamma^+$ is related to the impurity concentration $n_I$
by $\pi\Gamma^+ = n_I/[N(0)\pi]$. When $V_{imp}$ is very
small $c\to\infty$, we recover the Born limit in which case we
denote $\Gamma^+/c^2$ as $t^+$, and for $V_{imp}$ very large
$c\to 0$ which is called the unitary limit. A realistic case
would be for finite $c$. Fits to experimental data have given
values of the order of 0.2.\cite{schach1}  From the definition of $c$ and
taking $N(0)$ to be $1/W$ for
$c=0.2$, $V_{imp}\simeq 1.3\,W$, where $W$
is the band width.

\section{Numerical Results}

\begin{figure}
  \includegraphics[width=9cm]{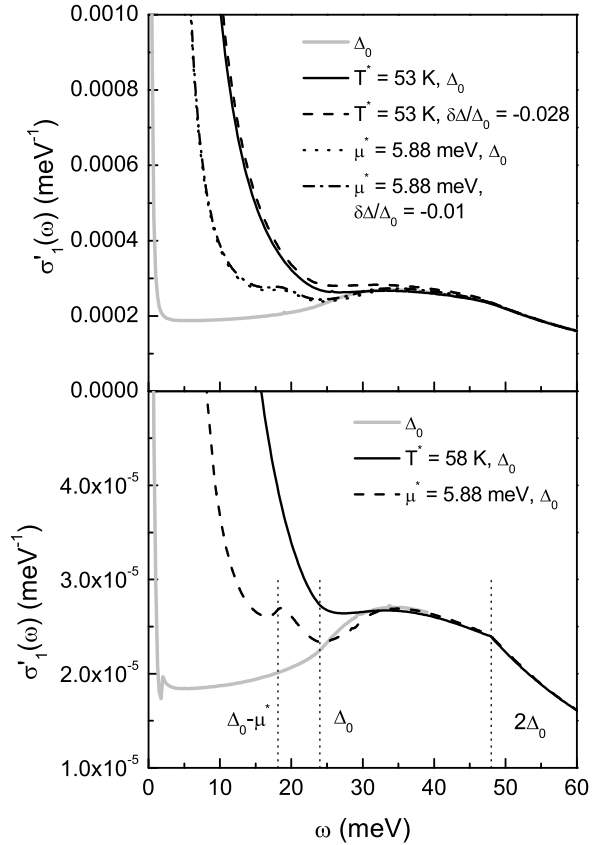}
  \caption{Top frame: the real part of the optical conductivity
    $\sigma'_1(\omega)$ as a function of frequency $\omega$.
    The solid gray is for the unirradiated equilibrium
    sample with temperature $T=3\,$K, gap $\Delta_0=24\,$meV,
    and an impurity content described by $t^+=0.1\,$meV.
    The solid (dashed) curve includes an excess quasiparticle density
    $n=0.03$ in the $T^\ast$-model ($T^\ast = 53\,$K) with the
    gap unaltered (altered to $0.972\Delta_0$). The dotted (dash-dotted)
    curve is for the $\mu^\ast$-model ($\mu^\ast = 5.88\,$meV)
    with the gap unaltered (altered to $0.99\Delta_0$). The
    bottom frame is the same as the top frame except that now
    the impurity content is greatly reduced to $t^+=0.01\,$meV,
    with a single $T^\ast=58\,$K (solid line) and a $\mu^\ast=5.88\,$meV
    (dashed line).
    }
  \label{fig:2}
\end{figure}

In the top frame of Fig.~\ref{fig:2} we show results for the
real part of the optical conductivity $\sigma'_1(\omega)$ as a
function of frequency $\omega$ in computer units. To get the
actual conductivity one must multiply the computer results by
$\Omega_p^2/4\pi$ with $\Omega_p$ the plasma frequency. The
parameters used for the run are $\Delta_0=24\,$meV, $T=3\,$K
which is low enough that it is representative of zero
temperature and $t^+=0.1\,$meV. The impurity scattering
is taken in the Born limit $(c\to\infty)$.
The solid gray curve labeled simply by $\Delta_0$ on the
figure is the conductivity in the equilibrium limit, i.e.
$n=0$, and is included for comparison. The other four curves
are with a finite non-equilibrium excess quasiparticle density
of $n=0.03$. This value of $n$ is large compared to a
value that one may realistically have in an experiment. The results
obtained, however, are illustrative of what one might expect
even for smaller $n$. For the $\mu^\ast$-model
$\mu^\ast\simeq 0.25\Delta_0$ and the change in the gap
$\delta\Delta\simeq -0.01\Delta_0$ which is very small. In
the $T^\ast$-model the corresponding number is
$T^\ast\simeq 0.19\Delta_0$ and $\delta\Delta = -0.028\Delta_0$,
again small but three times bigger than in the $\mu^\ast$ model.
In the top frame of Fig.~\ref{fig:2} the solid black and dashed curves are
for $T^\ast = 53\,$K with and without inclusion of the small
gap change. We see that the direct thermal effects of $T^\ast$
are much more important than any slight change in the
gap. This also holds for the $\mu^\ast$-model with dotted
(no gap change) and dash-dotted (with gap change) curves.
Comparison with the $n=0$ case shows that
the real part of the conductivity is strongly affected
by the inclusion of a chemical potential $\mu^\ast = 5.88\,$meV
and that there is significant difference between the predictions
of the $T^\ast$ and $\mu^\ast$ models. In particular, for the
$\mu^\ast$-model there is a distinct structure predicted to occur
at the value of the gap $\Delta_0$ minus the
chemical potential $\mu^\ast$. The characteristic
of these structures is more easily seen in the bottom frame
of Fig.~\ref{fig:2} where additional results for $\sigma'_1(\omega)$
vs $\omega$ are given for the same gap value of $24\,$meV but now the
impurity content is reduced by an order of magnitude to
$t^+=0.01\,$meV. The gray solid curve
which applies to the equilibrium case, shows a slight change
in slope at $\omega=\Delta_0=24\,$meV as well as at
$\omega = 2\Delta_0 = 48\,$meV. These structures are characteristic
of the logarithmic  van Hove singularity in the $d$-wave
quasiparticle equilibrium density of states. In comparison to the
gray solid curve, the dashed curve which has $\mu^\ast = 5.88\,$meV
shows an additional pronounced structure at 
$\sim 18\,$meV which is the value of $\Delta_0-\mu^\ast$ in this
case. When we also include a small
shift in the gap we do see it as a small shift in the
$2\Delta_0$ structure which corresponds
to the smallest optical frequency which can connect both logarithmic
van Hove singularities in initial and final electron state. By
contrast, the structure at $\Delta_0-\mu^\ast$ corresponds to a
transition between the top of the occupied region of the
density of states as the initial state with the van Hove singularity
as the final state. The remaining curve in the bottom frame
of Fig.~\ref{fig:2} is for the $T^\ast$ model (solid line).
Now the region below $2\Delta_0$ is quite smooth and shows no
discernible structure in contrast to the $\mu^\ast$-model.
This is expected since in the $T^\ast$-model the non-equilibrium
distribution is assumed to be thermal.

\begin{figure}[tp]
  \vspace*{-7mm}
  \includegraphics[width=9cm]{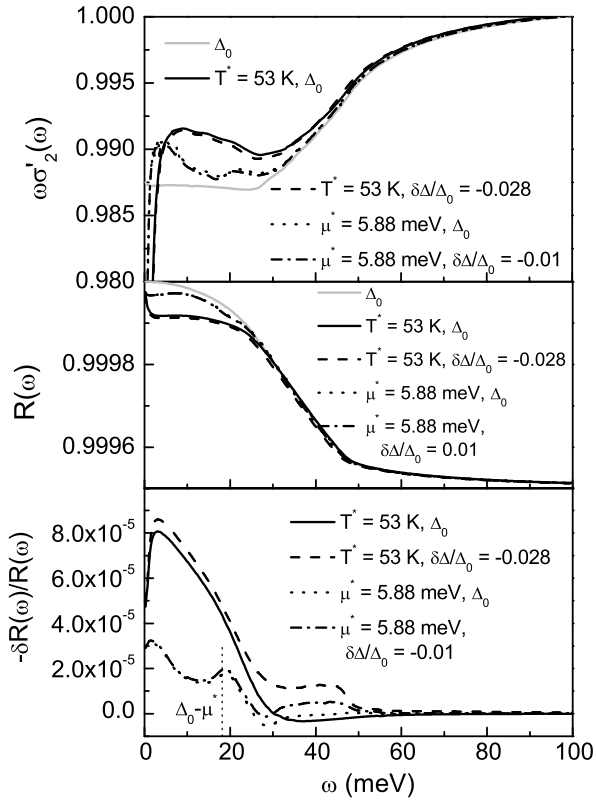}
  \caption{Top frame: same as in the top frame of Fig.~\protect{\ref{%
        fig:2}} except that now $\omega\sigma'_2(\omega)$ is plotted,
    i.e. $\omega$ times the imaginary part of the conductivity
    which is related
    to the inverse square of the penetration depth at $\omega\to0$.
    Middle frame: the reflectance as a function of $\omega$.
    Bottom frame: the normalized change in the reflectance
    $\delta R(\omega)/R(\omega)$ vs $\omega$. To get these we
    have used $\Omega_p = 2\,$eV and $\varepsilon_\infty = 1$.
    }
  \label{fig:3}
\end{figure}

Fig.~\ref{fig:3} shows additional results for the case $t^+=0.1\,$meV.
In the top frame we show $\omega\sigma'_2(\omega)$ 
($\omega$ multiplied by the imaginary part
of the conductivity) also denoted by a frequency dependent inverse
penetration depth $\lambda_L^{-2}(\omega,T)$ in the literature. The
middle frame shows the reflectance while the bottom frame shows
the normalized difference in reflectance between non-equilibrium
and equilibrium case, i.e. $\delta R(\omega)/R(\omega)$. This is
often the quantity that is measured directly in non-equilibrium
pump probe experiments and is included here for convenience.
Considering the top frame first, we see significant differences
in the value of $\omega\sigma'_2(\omega)$ for frequencies $\omega$
below the gap $\Delta_0=24\,$meV. In particular, a peak which does
not exist in the solid gray curve for the equilibrium case develops in
both, the $T^\ast$ and the $\mu^\ast$-model. The dotted and dash-dotted
curves apply to the $\mu^\ast$-model without and with change in
the gap included while the solid and dashed curves are for the
$T^\ast$-model. The theory predicts a sharper peak which forms
at smaller energy for $\mu^\ast$ as compared with the $T^\ast$-model.
The center frame of Fig.~\ref{fig:3} gives the reflectance
$R(\omega)$ vs $\omega$ for five  cases as in the other frames.
As in the top frame, the largest
differences in this set of curves occur at frequencies below
the gap $\omega < 24\,$meV. In that case the $T^\ast$-model
(solid black and dashed curves without and with the non-equilibrium
change in the gap included) predicts a very rapid drop in
$R(\omega)$ at the lowest values of $\omega$ and then a
plateau before a second rapid drop which sets in around the gap.
The curve for the $\mu^\ast$-model, dotted and dash-dotted without and
with gap change, is in comparison more structured and in fact
shows a small peak in the region 10 to $15\,$meV. The bottom
frame of Fig.~\ref{fig:3} serves to emphasize these differences
even more. What is plotted is the difference between
non-equilibrium and equilibrium value of $R(\omega)$ normalized to this
equilibrium value. The quantity $-\delta R(\omega)/R(\omega)$
shows a very rapid rise out of $\omega = 0$ which is larger for
the $T^\ast$ model than for the $\mu^\ast$-model (by roughly
a factor of four). After a maximum is reached, a minimum is
seen only in the $\mu^\ast$ case which shows a second maximum
within the gap region (corresponding to the structures in the
conductivity in the top frame of Fig.~\ref{fig:2} at $\Delta_0-\mu^\ast$)
before the change in reflectance becomes
quite small. In both models the region between $\Delta_0$ and
$2\Delta_0$ is also affected and is quite sensitive to changes
in the gap value.

\begin{figure}[tp]
  \includegraphics[width=9cm]{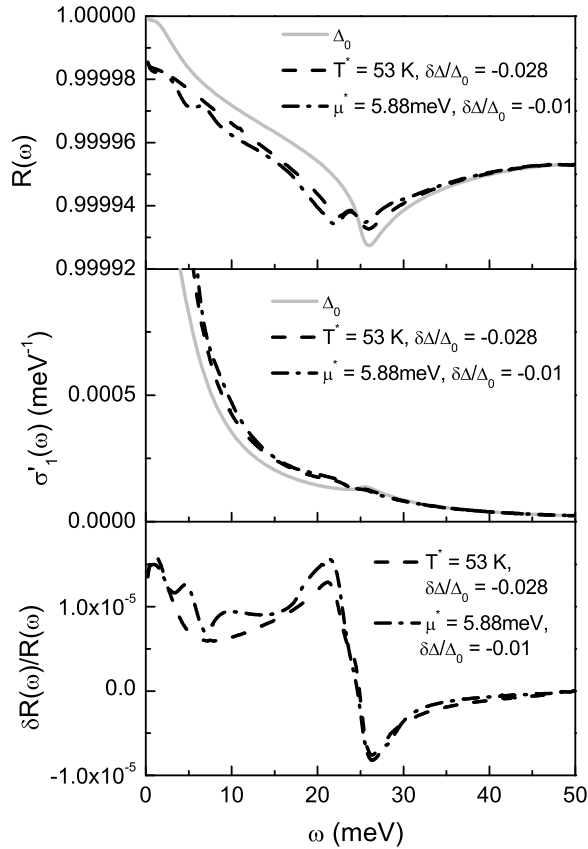}
  \caption{Reflectance $R(\omega)$ (top frame), real part of the optical
  conductivity $\sigma'_1(\omega)$ (middle frame), and normalized
  reflectance difference $\delta R(\omega)/R(\omega)$ (bottom frame)
  vs $\omega$ for a case with $\Delta_0 = 24\,$meV, $T=3\,$K,
  $\Gamma^+=0.01\,$meV, and $c=0.2$ near the unitary limit.
  The gray curve in each of the top two frames is the
  equilibrium case shown for comparison. The dashed curve is
  for the $T^\ast$-model and the dashed-dotted curve for the
  $\mu^\ast$-model.
    }
  \label{fig:4}
\end{figure}

We have carried out additional calculations for values of $c$ in
the impurity potential away from the Born limit. Here we report
only one case. In Fig.~\ref{fig:4} we show results
for $\Delta_0=24\,$meV, as in a previous figure with $T=3\,$K,
 $\Gamma^+=0.01\,$meV and $c=0.2$ which is near the unitary
limit. What is shown in the top frame is the
reflectance as a function of frequency. The gray curve is the
equivalent equilibrium case included for comparison. The
dashed curve is based on the $T^\ast$-model and the dash-dotted
curve on the $\mu^\ast$-model. The middle frame gives the
corresponding results for the real part of the optical
conductivity. Comparison of these results
with the equivalent results shown in the bottom frame of
Fig.~\ref{fig:2} reveales that the Born limit is a better case
in which to investigate non-equilibrium effects in the sense
that, for the same non-equilibrium quasiparticle density $n$
the effects are less pronounced for the $c=0.2$ case. In contrast
to the Born limit, the $\sigma'_1(\omega)$ vs $\omega$ (solid
gray curve) is now much closer to the non-equilibrium results
(dashed or dash-dotted curves). The normalized difference
$\delta R(\omega)/R(\omega)$ vs $\omega$ is shown in the
bottom frame of Fig.~\ref{fig:4}. The differences between the
$T^\ast$-model (dashed curve) and the $\mu^\ast$-model
(dash-dotted curve) are less pronounced than was found in
the bottom frame of Fig.~\ref{fig:3}.
 In our previous analysis of the optical
conductivity of various cuprates we have found, in several
cases, the need to include in our calculations some
impurity scattering in the unitary limit.\cite{schach9,schach2}
On the other hand, in some of the very pure samples grown in
BaZnO$_3$ crucibles the mean free path at low temperatures
is found to be of the order of a micron and a value of
$c\ne 0$ is needed.\cite{schach1}

\section{Simple analytic results}

It is possible and instructive to derive some simple analytical
results for the non-equilibrium conductivity calculated
numerically in the previous section. This can provide insight
into the physics involved as it makes explicit the dependence
on excess quasiparticle density $n$ of various quantities.
In terms of the London penetration depth $\lambda_L^{-2}(0)$
the penetration depth in the pure limit at finite temperature and finite
$\mu^\ast$ is given by
\begin{equation}
  \label{eq:6}
  \lambda_L^{-2}(T,\mu^\ast) =
  \lambda_L^{-2}(0)\left[1+\int\!dE\,N(E)\frac{\partial
      f_T(E-\mu^\ast)}{\partial E}\right],
\end{equation}
where $N(E)$ is the quasiparticle density of states. For small
$\mu^\ast$ at zero temperature the thermal factors in Eq.~(\ref{%
eq:6}) become a $\delta$-function; we can also
use the nodal approximation for $N(E) = E/\Delta_0$ and
get immediately
\begin{equation}
  \label{eq:7}
  \lambda_L^{-2}(0,\mu^\ast) =
  \lambda_L^{-2}(0)\left[1-\frac{\mu^\ast}{\Delta_0}\right]
  = \lambda_L^{-2}(0)\left[1-\sqrt{2n}\right].
\end{equation}
The London penetration depth at zero temperature is increased
by the presence of a non-equilibrium distribution and the
reduction in superfluid density follows a square root of $n$
law. This is also seen in the results presented in the top frame
of Fig. 3. A similar result also holds
for the $T^\ast$ model. From Eq.~(\ref{eq:6}) with $\mu^\ast=0$
and $T$ replaced by $T^\ast$ we get
\begin{eqnarray}
  \lambda_L^{-2}(T^\ast,\mu^\ast) &=&
  \lambda_L^{-2}(0)\left[1-\frac{2\ln 2}{\Delta_0} T^\ast
    \right]\nonumber\\
  & =& \lambda_L^{-2}(0)\left[1-
    \frac{2\ln(2)\sqrt{12 n}}{\pi}
        \right].
  \label{eq:8}
\end{eqnarray}
The intermediate formula in (\ref{eq:8}) is just the well known linear
in $T$ law for a pure $d$-wave superconductor. The coefficient
of the $\sqrt{n}$ term in the last expression in (\ref{eq:8})
is $1.53$ while in the $\mu^\ast$-model the coefficient was
$1.4$. For the same non-equilibrium density $n$ the superfluid
density is less reduced from its equilibrium value in the
$\mu^\ast$-model. This mirrors but is the opposite of what was
found for the superconducting gap. In both models the reduction
follows a $\sqrt{n^3}$ law with coefficients 1.9 in the
$\mu^\ast$-model as compared with 5.4 in the $T^\ast$-model.
The physics of the reduction in this case is that the states
around the Fermi surface occupied by the excess quasiparticles
more effectively block the formation of the condensate in the
$\mu^\ast$-model than they do in the $T^\ast$-model where they are
distributed over higher energy states. Formulas including finite
temperature corrections are derived in the Appendix.

In the limit of weak scattering, i.e. $\Gamma^+\to 0$
(temperature dominated regime), self
consistency is not required in Eq.~(\ref{eq:5}) where
$\tilde{\omega}$ on the right hand side can simply be replaced
by $\omega$. This corresponds to the temperature dominated
regime with $\gamma\ll T$. Here $\gamma$ is the impurity
scattering rate in the superconducting state and is ferquency
dependent. Considerable simplifications result as seen
in the work of Hirschfeld {\it et al.}\cite{hirschf} Generalizing
their result for $0<\gamma\ll T$ to the non-equilibrium case,
the real part of the conductivity is given by
\begin{eqnarray}
  \sigma_1(T,\omega)&\simeq&\frac{\Omega_p^2}{4\pi}
  \int\limits_{-\infty}^\infty\!d\nu\,\left(-\frac{\partial
      f_T(\nu-\mu^\ast)}{\partial \nu}\right)N(\nu)\nonumber\\
  &&\times
  \Im{\rm m}\frac{1}{\omega-i\tau^{-1}(\nu)}
  \label{eq:10}
\end{eqnarray}
where the scattering rate $\tau^{-1}(\omega) = \gamma(\omega) =
-\Im{\rm m}\,\tilde{\omega}(\omega+i0^+)$ is to be determined
by Eq.~(\ref{eq:5}) with $\tilde{\omega}$ replaced by $\omega$
on the right hand side. The result for a general value of $c$ is
\begin{equation}
  \label{eq:11}
  \tau^{-1}(\omega) = \pi\Gamma^+\frac{\omega}{\Delta_0}
  \frac{c^2+A_+(\omega)}{c^4+2c^2A_-(\omega)+A^2_+(\omega)},
\end{equation}
where
\begin{equation}
  \label{eq:11a}
  A_\pm(\omega) = \left(\frac{2\omega}{\pi\Delta_0}\right)^2
  \left[\left(\frac{\pi}{2}\right)^2\pm\ln^2\left(\frac{2\Delta_0}
   {\omega}\right)\right].
\end{equation}
Two limits are normally considered. The Born limit with
$c\to\infty$ and the unitary limit with $c=0$. In these
limits we get:
\begin{equation}
  \label{eq:11b}
  \tau^{-1}(\omega) = \left\{\begin{array}{lcl}
   \frac{\pi^3\Gamma^+}{4}\frac{\Delta_0}{\omega}
   \frac{1}{(\pi/2)^2+\ln^2(2\Delta_0/\omega)}&\quad&c=0\\
   \frac{\pi\Gamma^+}{c^2}\frac{\omega}{\Delta_0}&\quad&
   c\to\infty.
   \end{array}\right .
\end{equation}
In this last equation we can write $\pi\Gamma^+ = n_I/[\pi N(0)]$
and making use of the relation $c^{-1} = 2\pi N(0)V_{imp}$, we get
$\pi\Gamma^+/c^2 = 2\Gamma_B$ with $\Gamma_B = 2\pi n_I N(0)V_{imp}^2$,
the well known formula for impurity scattering (Fermi Golden Rule).
We can now use the scattering
rates in the conductivity formula (\ref{eq:10}).

We begin with the $\mu^\ast$-model. For small temperatures $\ll
\mu^\ast$, $(\gamma\ll T \ll \mu^\ast)$
\begin{equation}
  \label{eq:13}
  \sigma_1(T,\omega) = \frac{\Omega_p^2}{4\pi}\frac{\mu^\ast}{\Delta_0}
     \frac{\tau(\mu^\ast)}{1+\omega^2\tau^2(\mu^\ast)}.
\end{equation}
The finite frequency conductivity samples only the scattering
time at the single frequency $\nu=\mu^\ast$ in this model. They
are
\begin{equation}
  \label{eq:13a}
  \tau^{-1}(\mu^\ast) = \left\{\begin{array}{lcl}
\frac{\pi^3\Gamma^+}{4}\frac{1}{\sqrt{2n}}
  \frac{1}{(\pi/2)^2+\ln^2(\sqrt{2/n})},&\quad& c=0\\
  2\Gamma_B\sqrt{2n},&\quad&c\to\infty.
  \end{array}\right.
\end{equation}
We see that in the Born limit $\tau^{-1}(\mu^\ast)$ is directly
proportional to the square root of the excess quasiparticle
density, while it varies as its inverse in the unitary limit with
some other additional, weaker logarithmic dependence on
the square root of $n$. This means that the
half width of the microwave conductivity will be strongly affected
by changes in $n$ and that the effects will be opposite in
Born and unitary limit. These features are directly related to the
details of the quasiparticle density of states at small $\omega$
in a $d$-wave superconductor and $\tau^{-1}(\mu^\ast)$ as a
function of $\mu^\ast$ gives this information directly.

From Eq.~(\ref{eq:13}) one can easily compute the spectral optical
weight remaining under $\sigma_1$. We define $A$ as
\begin{equation}
  A = \int\limits_0^\infty\!d\omega\,\sigma_1(\omega) =
   \frac{\Omega_p^2}{8}\frac{\mu^\ast}{\Delta_0}
   = \frac{\Omega_p^2}{8}\sqrt{2n},
  \label{eq:14}
\end{equation}
which gives a square root of $n$ law independent of Born or unitary
limit and of temperature, but the restriction $\gamma\ll T\ll\mu^\ast$ must
be noted. This accounts exactly for the
missing superfluid density brought about by the excess
quasiparticles.

\begin{figure}[tp]
  \includegraphics[width=9cm]{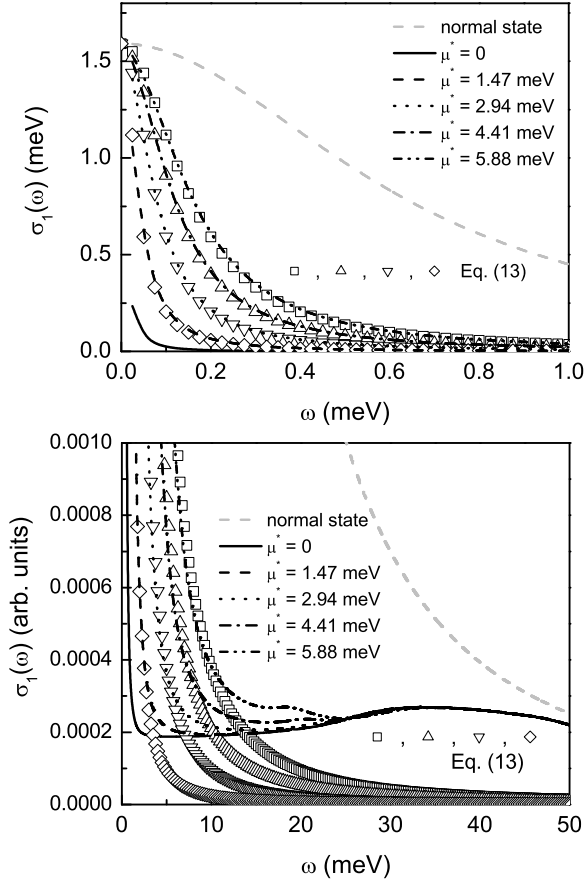}
  \caption{Comparison of exact results (black lines) for $\sigma_1(T,\omega)$
[Eq.~\protect{\eqref{eq:2}}] with the approximate analytic result of
Eq.~\protect{\eqref{eq:13}} (open symbols).
The top frame is for frequencies
restricted to $0\le\omega\le 1\,$meV while the bottom frame spans to
$50\,$meV. The gray
dashed curve is the normal state for $t^+=0.1\,$meV and $T=3\,$K.
The superconducting gap is
$24\,$meV, five values of $\mu^\ast$ are shown.}
\label{fig:5}
\end{figure}

In the top frame of Fig.~\ref{fig:5} we compare results based on
the approximate Eq.~\eqref{eq:13} with full numerical results based
on the exact
expression \eqref{eq:2}. The same impurity parameters as presented
in Fig.~\ref{fig:2} are used, namely $t^+ = 0.1\,$meV with a gap
$\Delta_0 = 24\,$meV and $T=3\,$K. Several
values of the non-equilibrium chemical potential $\mu^\ast$ are
considered: $\mu^\ast=0$ (black), $\mu^\ast=1.47\,$meV (dashed),
$\mu^\ast=2.94\,$meV (dotted), $\mu^\ast=4.41\,$meV (dash-dotted),
and $\mu^\ast=5.88\,$meV (dash-double dotted). The black curves are results
based on Eq.~\eqref{eq:2} while the open symbols are based
on Eq.~\eqref{eq:13}. The frequency $\omega$ is restricted to
small values below $1.0\,$meV. We see remarkable agreement
between the full theory and Eq.~\eqref{eq:13}. This agreement, of
course, will fail as $\omega$ is increased outside the validity of
the approximation used to derive Eq.~\eqref{eq:13}. This is seen
in the bottom frame of Fig.~\ref{fig:5} where $\omega$ now
goes up to $50\,$meV, i.e. up to about twice the gap value.
For this higher frequency region no simple analytic result can be
obtained and it is necessary to return to the numerical results
of the previous section.

\begin{figure}[tp]
  \includegraphics[width=9cm]{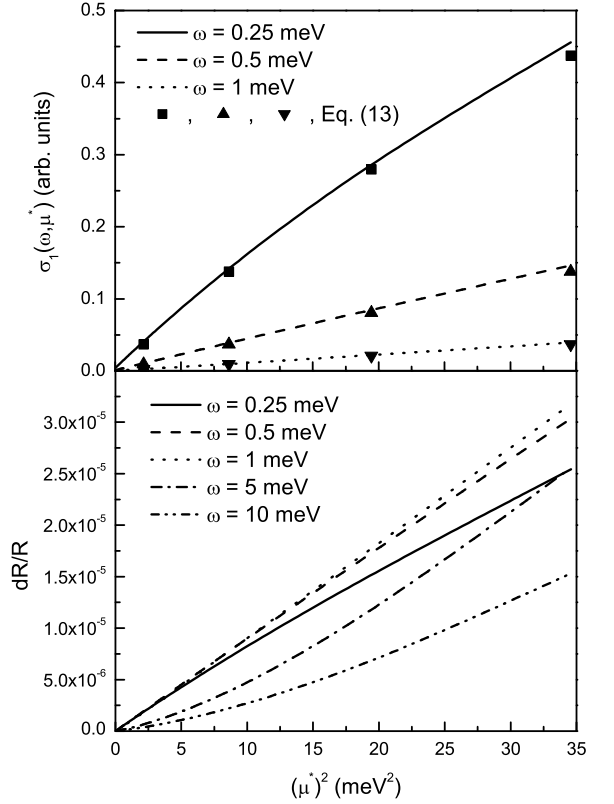}
\caption{Top frame: Real part of the conductivity $\sigma_1(\omega,%
\mu^\ast)$ at three frequencies $\omega = 0.25$, $0.5$, and
$1.0\,$meV as a function of the square of the chemical potential
$\mu^\ast$ $(\sim n)$. The lines are based on exact numerical evaluations
of Eq.~\protect{\eqref{eq:2}} while the solid symbols are based
on the approximate analytic formula \protect{\eqref{eq:13}}.
Bottom frame: The normalized change in reflectivity
$\delta R/R$ at five frequencies, $\omega=0.25$, $0.5$, $1.0$,
$5.0$, and $10.0\,$meV as a function of $(\mu^\ast)^2$ $(\sim n)$
based on numerical evaluation of Eq.~\protect{\eqref{eq:2}}. In
both frames $\Delta_0=24\,$meV, $t^+=0.1\,$meV, and $T=3\,$K.}
\label{fig:6}
\end{figure}

So far, except for Fig.~\ref{fig:5}, we have emphasized the expected
frequency dependence of the optical parameters for a given value
of excess quasiparticle density $n$. The data in Fig.~\ref{fig:5}
can be replotted to give the conductivity $\sigma_1(\omega,\mu^\ast)$
at a given frequency $\omega$ as a function of $\mu^\ast$. It is
clear from Eqs.~\eqref{eq:13} and \eqref{eq:13a} that in the
Born limit for $\omega\tau(\mu)\gg 1$, $\sigma_1(T,\omega)$ will
be proportional to $(\mu^\ast)^2$ and inversely proportional to
$\omega^2$. This is shown in the top frame of Fig.~\ref{fig:6}
where exact numerical results (solid lines) are compared with
the approximate results of Eq.~\eqref{eq:13} (solid symbols). The
expected linear
variation with the excess quasiparticle density
$[n\propto(\mu^\ast)^2]$ is pretty well verified.
Since in experiments it is the change in reflectance
$R$ which is measured directly we show in the bottom frame of
Fig.\ref{fig:6} results of this quantity at five frequencies, again as
a function of $n\propto(\mu^\ast)^2$. These data, based on exact
numerical calculations, show that the dependence of $\delta R/R$
on $n$ is not exactly linear.

Similar results to those embodied in Eqs.~\eqref{eq:10} to
\eqref{eq:14} for the $\mu^\ast$-model
can be obtained in the $T^\ast$ model as well. 
We give only a few analytic results here without comparison to
full numerical work. First we note that the residual absorption
is
\begin{eqnarray}
  A &=& \frac{\Omega_p^2}{8}\int\!d\omega\,\left(-\frac{\partial
      f_T(\omega)}{\partial\omega}\right)N(\omega) =
  \frac{\Omega_p^2}{8}\frac{2\ln(2)}{\Delta_0} T^\ast\nonumber\\
 &\simeq&  1.08\,\frac{\Omega_p^2}{8}\sqrt{2n},
  \label{eq:15}
\end{eqnarray}
which is just the missing superfluid density due to $T^\ast$
[see Eq.~\eqref{eq:6}]. The situation is, however, more
complicated when the details of the microwave conductivity are
considered. A simple form such as Eq.~(\ref{eq:13}) can no longer
be derived and we must return to Eq.~(\ref{eq:10}) which gives
\begin{equation}
  \label{eq:16}
  \sigma_1(\omega) \simeq \frac{\Omega_p^2}{4\pi}
  \int\limits_{-\infty}^\infty\!d\nu\,\left(-\frac{\partial
      f_{T^\ast}(\nu)}{\partial\nu}\right)\frac{\vert\nu\vert}{\Delta_0}
      \frac{\tau(\nu)}{1+\omega^2\tau^2(\nu)},
\end{equation}
where the temperature in the Fermi function is $T^\ast$ rather
than $T$. Even if the frequency $\nu$ in the logarithm
that occurs in the Eqs.~\eqref{eq:11} for $\tau(\nu)$
is replaced by $T^\ast$, the integral in Eq.~(\ref{eq:16})
cannot be done analytically for a finite value of $\omega$.
In effect several values of $\tau(\omega)$ are sampled in
Eq.~(\ref{eq:16}) in contrast to the $\mu^\ast$-model where
only $\tau(\mu^\ast)$ enters. This is a true difference
between the two models. If $\omega = 0$, however, we do get
an analytic result (assuming $\gamma\ll T\ll T^\ast$):
\begin{subequations}
\begin{equation}
  \label{eq:17a}
  \sigma(\omega\to0) = \left\{\begin{array}{lcl} 
  \frac{\Omega_p^2}{4\pi}\frac{1}{2\Gamma_B}, && c\to\infty,\\
   \frac{\Omega_p^2}{4\pi}\frac{16 n}{\pi^3\Gamma^+}\left[
  \left(\frac{\pi}{2}\right)^2+\ln^2\left(\frac{\pi}{\sqrt{3n}}
  \right)\right],&& c=0.
  \end{array}\right.
\end{equation}
In the opposite limit, $\omega$ going to $\infty$
\begin{equation}
  \label{eq:17b}
  \sigma(\omega) = \left\{\begin{array}{lcl} 
   \frac{\Omega_p^2}{4\pi} 8n\frac{\Gamma_B}{\omega^2},&&
  c\to\infty,\\
   \frac{\Omega_p^2}{4\pi}\pi^3\frac{\Gamma^+}{4\omega^2}
   \frac{1}{(\pi/2)^2+\ln^2(\pi/\sqrt{3n})},&&c=0.
   \end{array} \right.
\end{equation}
\end{subequations}
We note that, in both cases, the dependence of the conductivity
on the excess quasiparticle density $n$ is very different in the
two impurity limits considered. The nature of the impurity scattering
is having a profound effect on the optical conductivity. For
$\omega\to 0$ in the Born limit, $n$ drops out of the expression
for the conductivity which then takes on its normal state value.
For the unitary limit, instead, it goes like
$n\ln^2(n)$. For $\omega\to\infty$ in the Born limit the
conductivity is proportional to $n$ and in the unitary limit
it varies inversely as $\ln^2(n)$. Similar results hold for the
$\mu^\ast$-model; only the numerical factors are changed as can
easily be verified from the general case \eqref{eq:13}. In
particular, the $c\to\infty$ result of \eqref{eq:17a} stays
unchanged and the $c=0$ result is reduced by a factor of two,
and the numerical factor in the logarithm $\pi/\sqrt{3}=1.81$
is reduced by $\sqrt{2}=1.41$; the $\omega\to\infty$ result
of \eqref{eq:17b} is reduced by two and for $c=0$ we get the same
result except for the numerical factor in the logarithm which
goes from 1.81 to 1.41. While there are some quantitative
differences between the results for the real part of the
conductivity between $T^\ast$ and $\mu^\ast$-models, in the
restricted parameter range in which our analytic results apply,
and for $\omega\to\infty$ the excess quasiparticles dominate
the value of the conductivity. The condition
$\gamma\ll T\ll T^\ast$ implies that
$\Omega_p^2/(4\pi)(T^\ast/\Delta_0)^2(2/3)\pi^2(\Gamma_B/\omega^2)$
[which is the result for the Born limit in Eq.~\eqref{eq:17b}] is
much larger than its value when $n=0$, in which case $T$ replaces
$T^\ast$ in this expression; we note that the condition
$T\ll T^\ast$ has been assumed. The same holds for the unitary limit.

Finally, we consider possible modifications of the so-called universal
limit. Applying a nodal approximation to the formula for the
$T<\Delta_0$ conductivity, Eq.~(\ref{eq:2}), it reduces at
zero frequency\cite{wk}
\begin{eqnarray}
  \sigma_1(T,0) &=& \frac{\Omega^2_p}{4\pi}\frac{1}{\pi\Delta_0}
    \int\limits_{-\infty}^\infty
        \!d\omega\,\left(-\frac{\partial f_T(\omega)}{\partial\omega}
        \right)\nonumber\\
  &&\times\left[1+\frac{\omega}{\gamma(\omega)}\tan^{-1}
        \left(\frac{\omega}{\gamma(\omega)}\right)\right],
  \label{eq:18}
\end{eqnarray}
for $T\to 0$ this gives $\sigma_1(0,0) = [\Omega^2_p/(4\pi)][1/%
(\pi\Delta_0)] \equiv\sigma_{00}$ which is the well known
universal limit\cite{lee,wu} independent of impurity scattering.
Generalization to non-equilibrium in the $\mu^\ast$-model gives
\begin{equation}
  \label{eq:19}
  \sigma_1(0,0) = \frac{\Omega^2_p}{4\pi}\frac{1}{\pi\Delta_0}
  \left[1+\frac{\mu^\ast}{\gamma(\mu^\ast)}\tan^{-1}
    \left(\frac{\mu^\ast}{\gamma(\mu^\ast)}\right)\right],
\end{equation}
where $\gamma(\omega)\simeq\gamma[1+b(\omega/\gamma)^2]$
(with $b$ a constant\cite{hirschf})
obtained from a self consistent solution of Eq.~(\ref{eq:5})
for small but finite $\omega$.
For $\mu^\ast\ll\gamma$ we can replace $\gamma(\mu^\ast)$
by its constant value of $\gamma$ and find
\begin{equation}
  \label{eq:20}
  \sigma_1(0,0) = \sigma_{00}\left[1+\left(\frac{\mu^\ast}
      {\gamma}\right)^2\right] = \sigma_{00}\left[1+
      \left(\frac{\Delta_0}{\gamma}\right)^2 2n\right].
\end{equation}
The connection to $\sigma_{00}$ depends linearly on $n$ and
on the inverse square of $\gamma$. Universality is lost.

\section{Conclusion}

We have presented results for the modifications of the optical
conductivity $\sigma_1(T,\omega)$ that are brought about by
introduction of a finite non-equilibrium excess quasiparticle
density $n$. To describe the non-equilibrium distribution in
energy, we use two simplified models that have been found to be
useful for the $s$-wave case. While not expected
to be accurate the $\mu^\ast$-model of Owen and
Scalapino\cite{owen} and the $T^\ast$-model of Parker\cite{parker1,%
  parker2} have the great advantage that the physics involved
becomes transparent. The frequency dependent conductivity shows
distinct features associated with the chemical potential which
should be observable. While both, the $\mu^\ast$ and the
$T^\ast$-model give similar results qualitatively, there are
some important quantitative differences. Numerical results
for the real and the imaginary part of $\sigma$ are given
separately, as is the reflectance and the normalized change
in reflectance brought about by the excess quasiparticle
density. We have also given several analytic results which we
hope will prove helpful in the analysis of experimental data.

\section*{Acknowledgment}
 
Research supported by the Natural Sciences and Engineering
Research Council of Canada (NSERC) and by the Canadian
Institute for Advanced Research (CIAR).

\appendix
\section{Finite temperature corrections}

We consider finite temperatures in the penetration depth.
The algebra is simplest
in the $T^\ast$-model. The thermal quasiparticle density in
a $d$-wave superconductor at temperature $T$ is given by
$(\pi^2/12)(T/\Delta_0)^2$ and so the excess non-equilibrium
density $n$ in the $T^\ast$-model is
\begin{equation}
  \label{eq:1s}
  n = \frac{\pi^2}{12}\left[\left(\frac{T^\ast}{\Delta_0}\right)^2
  -\left(\frac{T}{\Delta_0}\right)\right].
\end{equation}
Thus, we have
\begin{equation}
  \label{eq:2s}
  \frac{T^\ast}{\Delta_0} = \sqrt{\frac{12 n}{\pi^2}+
  \left(\frac{T}{\Delta_0}\right)^2},
\end{equation}
which reduces to the known result $T^\ast/\Delta_0 =%
2\sqrt{3n}/\pi$ for $T=0$.

It is easy to work out the small $T$ correction
\begin{equation}
  \label{eq:3s}
  \frac{T^\ast}{\Delta_0} = \frac{2\sqrt{3}}{\pi}\sqrt{n}
  +\frac{\sqrt{3}}{12}\frac{\pi}{\sqrt{n}}\left(\frac{T}{\Delta_0}
  \right)^2,
\end{equation}
while the small $n$ correction to finite $T$ is
\begin{equation}
  \label{eq:4s}
  \frac{T^\ast}{\Delta_0} = \frac{T}{\Delta_0}+
  \frac{6n}{\pi^2}\frac{\Delta_0}{T}.
\end{equation}

Equation (\ref{eq:3s}) is valid for $12n/\pi^2 > (T/\Delta_0)^2$
and (\ref{eq:4s}) for $12n/\pi^2 < (T/\Delta_0)^2$. In
(\ref{eq:3s}) the correction goes like $T^2$ and is inversely
dependent on $\sqrt{n}$ while in (\ref{eq:4s}) it is linear in
$n$ and inversely dependent on temperature. The reduction in
gap value due to the non-equilibrium density $n$ also has a
temperature dependence which is
\begin{equation}
  \label{eq:5s}
  \frac{\delta\Delta(n)}{\Delta_0} =
  -4\left\{\left[\frac{12n}{\pi^2}+\left(\frac{T}{\Delta_0}
   \right)^2\right]^{3/2}-\left(\frac{T}{\Delta_0}\right)^3
  \right\},
\end{equation}
which reduces to the known result $-32\sqrt{(3n)^3}/\pi^3$
at $T=0$. It is easy to work out the lowest order $T$ correction
to this result as well as at finite $T$, the lowest order $n$
correction. As these corrections come in as higher order
effects in the penetration depth we will not give the
expressions here.

The penetration depth follows directly from Eq.~(\ref{eq:8})
with $T^\ast$ replaced by Eq.~(\ref{eq:2s}). This gives
\begin{equation}
  \label{eq:6s}
  \frac{1}{\lambda_L^2(T^\ast,T)} =
  \frac{1}{\lambda_L^2(0)}\left[1-2\ln(2)\sqrt{\frac{12n}{\pi^2}
  +\left(\frac{T}{\Delta_0}\right)^2}\right].
\end{equation}
For $T/\Delta_0 < \sqrt{12 n/\pi^2}$ (small
$T$ correction) we get the first correction
for temperature
\begin{equation}
  \label{eq:7s}
  \frac{1}{\lambda_L^2(T^\ast,T)} =
  \frac{1}{\lambda_L^2(0)}\left[1-\frac{\ln(2)}{\pi}\sqrt{3n}-
   \frac{\pi\ln(2)}{2\sqrt{3n}}\left(\frac{T}{\Delta_0}\right)^2
   \right],
\end{equation}
and for $T/\Delta_0 > \sqrt{12 n/\pi^2}$ (small
$n$ correction) the first correction
for non-equilibrium to temperature $T$ is
\begin{equation}
  \label{eq:8s}
   \frac{1}{\lambda_L^2(T^\ast,T)} =
  \frac{1}{\lambda_L^2(0)}\left[1-2\ln(2)\frac{T}{\Delta_0}-
  \frac{12\ln(2)}{\pi^2}n\frac{\Delta_0}{T}\right],
\end{equation}
which follows directly from Eqs.~(\ref{eq:3s}) and (\ref{eq:4s}).
We see from Eq.~(\ref{eq:7s}) that the low temperature
behavior of the penetration depth has changed from a $T$ to
a $T^2$ law through the presence of the non-equilibrium density
$n$. This is analogous to the effect of impurities which also
bring about a $T$ to $T^2$ law change at low $T$. Of course,
at higher values of $T$ we recover a linear in $T$ law as is
indicated in Eq.~(\ref{eq:8s}) with a small correction for the
non-equilibrium density $n$.

A similar situation holds in the $\mu^\ast$-model but the
algebra is not as tidy and we will only give two results.
It is easy to work out the finite temperature correction to the
chemical potential. From its definition (\ref{eq:1s})
\begin{eqnarray}
  n\Delta_0 &=& \int\limits_0^{\mu^\ast}\!dE\,\frac{E}{\Delta(0)}
  f_T(E-\mu^\ast)+\int\limits_0^\infty\!dE\,f_T(E)\nonumber\\
  &=& \frac{{\mu^\ast}^2}{2\Delta_0}+\frac{\mu^\ast}{\Delta_0}
     T2\ln(2).
  \label{eq:9s}  
\end{eqnarray}
The solution for $\mu^\ast/\Delta_0$ is
\begin{equation}
  \label{eq:10s}
  \frac{\mu^\ast}{\Delta_0} = -2\ln(2)\,\frac{T}{\Delta_0}
  +\sqrt{2n+\left(\frac{2T\ln(2)}{\Delta_0}\right)^2},
\end{equation}
which properly reduces to $\sqrt{2n}$ at $T=0$ and to zero
for $n=0$. For $\mu^\ast/T\to\infty$ the expression for
$\lambda_L^{-2}$ remains that given Eq.~(\ref{eq:7}), and
we get
\begin{equation}
  \label{eq:12s}
\frac{1}{\lambda_L^2(T^\ast,T)} =
  \frac{1}{\lambda_L^2(0)}\left[1-
  \sqrt{2n+\left(\frac{2T\ln(2)}{\Delta_0}\right)^2}\right],  
\end{equation}
and the square bracket reduces to
$1-\sqrt{2n}$ at $T=0$ and shows a positive leading,
linear in temperature correction with the slope half the value
it would have in the equilibrium case. This is different from
the $T^\ast$-model and is due to the very different non-equilibrium
thermal distribution employed in the models which shows up at
very small $T$. At large $T$, the $\mu^\ast$-model gives
\begin{equation}
  \label{eq:13s}
  \frac{1}{\lambda_L^2(T^\ast,T)} =
  \frac{1}{\lambda_L^2(0)}\left[1-2\ln(2)\frac{T}{\Delta_0}+
  \frac{n}{2T}\Delta_0\right],
\end{equation}
which is the same as Eq.~(\ref{eq:8s}) except for a slightly
different numerical coefficient in the last term 0.5 rather
than 0.8.

\end{document}